\documentclass[11pt]{article}
\usepackage[body={16.6cm,23cm}]{geometry}

\usepackage[hang,small,center]{caption}
\usepackage{amsmath,amssymb}
\usepackage{amsfonts}
\usepackage{mathrsfs}
\usepackage{dsfont}

\usepackage{epic}
\usepackage{eepic}
\usepackage{graphicx}

\usepackage{pgf}
\usepackage{tikz}

\usepackage{color}

\usepackage{stmaryrd}

\usepackage{mathtools}
\DeclarePairedDelimiter{\floor}{\lfloor}{\rfloor}

\usepackage{cite}
\usepackage{enumerate}
\usepackage[curve]{xy}
\newcommand{\ds}{\displaystyle}

\renewcommand{\author}[1]{\large\rm #1\\ \bigskip}
\newcommand{\address}[1]{{\normalsize\it #1\\}\bigskip}
\renewcommand{\title}[1]{\bigskip\bigskip\Large\bf #1\bigskip\bigskip\\}

\newcommand{\Bigpsi}[3]{\phantom{\Psi}_2 \kern -.05em
\Psi_2\left(\genfrac{}{}{0pt}{}{#1}{#2}\biggl|#3\right)}

\newcommand{\bea}{\begin{eqnarray}}
\newcommand{\eea}{\end{eqnarray}}

\newcommand{\beq}{\begin{equation}}
\newcommand{\eeq}{\end{equation}}

\newcommand{\ii}{\mathsf{i}}

\newcommand{\ow}{\overline{\mathcal W}}
\newcommand{\w}{{\mathcal W}}
\newcommand{\s}{{\mathcal S}}

\newcommand{\cpar}{{\eta}}

\newcommand{\q}{{\mathsf q}}
\newcommand{\p}{{\mathsf p}}

\newcommand{\iW}{\mathcal{W}}
\newcommand{\iS}{\mathcal{S}}

\def\EXP{\textrm{{\large e}}}

\def\re{\mathop{\hbox{\rm Re}}\nolimits}
\def\im{\mathop{\hbox{\rm Im}}\nolimits}

\newcommand{\url}[1]{}

\renewcommand{\textcolor}[1]{}

\newcounter{app}
\newcounter{sapp}[app]
\def\theapp{\Alph{app}}
\newcommand{\app}[1]{
\refstepcounter{app}{\vspace{7mm}
\noindent\Large\bf Appendix
\theapp.
 \ #1 \par \vspace{5mm}}
\setcounter{equation}{0}
\def\theequation{\Alph{app}.\arabic{equation}}}

\begin{document}

\vglue 2cm

\begin{center}

\title{New solutions of the star-triangle relation with discrete and continuous spin variables}
\author{Andrew P.~Kels}
\address{Institut f\"{u}r Mathematik, MA 8-4, Technische Universit\"{a}t Berlin,\\
 Str. des 17. Juni 136, 10623 Berlin, Germany.}

\end{center}

\begin{abstract}

A new solution to the star-triangle relation is given, for an Ising type model of interacting spins containing integer and real valued components.  Boltzmann weights of the model are given in terms of the lens elliptic gamma function, and are based on Yamazaki's recently obtained solution of the star-star relation.  The star-triangle relation given here, implies Seiberg duality for the $4\!-\!d$ $\mathcal{N}=1$ $S_1\times S_3/\mathbb{Z}_r$ index of the $SU(2)$ quiver gauge theory, and the corresponding two component spin case of the star-star relation of Yamazaki.  A proof of the star-triangle relation is given, resulting in a new elliptic hypergeometric summation/integration identity.  The star-triangle relation in this paper contains the master solution of Bazhanov and Sergeev as a special case.  Two other limiting cases are considered one of which gives a new star-triangle relation in terms of ratios of infinite $q$-products, while the other case gives a new way of deriving a star-triangle relation that was previously obtained by the author.

\end{abstract}


\newpage


\section{Introduction}
The star-triangle relation is a distinguished form of the Yang-Baxter equation for Ising-type models on two-dimensional lattices.  In these
models the fluctuating variables, or ``spins'', are assigned to lattice
sites, and two spins interact only if they are connected by an edge of
the lattice.  Remarkably, many physically interesting models in this class can be solved exactly, for instance, the $2\!-\!d$ Ising \cite{Bax82}, and chiral Potts \cite{AuY87,Baxter:1987eq} models, and some others \cite{Zam-fish,FZ82,Kashiwara:1986,FV95,BMS07a,BMS07b,Bazhanov:2010kz} (see also \cite{BKS2,Bax02rip} for a review of other known cases).  The star-triangle relation plays the role of the integrability condition for these models.

Recently Bazhanov and Sergeev (BS) obtained an important ``master'' solution \cite{Bazhanov:2010kz} of the star-triangle relation, which contained all previously known solutions of this relation as particular cases, and provides interesting new examples.  The above master solution is expressed in terms of the elliptic gamma function, which contains two arbitrary free parameters $\p$ and $\q$, that play the role of elliptic nomes.  The spin variables for the corresponding statistical mechanical model take continuous real values on the circle.

Considered as a mathematical identity the BS master solution is identical to the elliptic beta integral of Spiridonov \cite{Spiridonov-beta}.  The latter discovery was central to the modern development of the theory of elliptic hypergeometric functions \cite{Spiridonov-essays}, and some recent works further highlight that some of these identities are connected to the integrability of lattice models of statistical mechanics.  Some examples include an extension of the BS master solution to the case of multi-component spins \cite{BS11,BKS}, and remarkable correspondences to Seiberg duality in supersymmetric gauge theories \cite{DolanOsborn,Spiridonov-statmech,Yamazaki2012, Yamazaki2013}.  Recently Yamazaki introduced a new integrable model \cite{Yamazaki2013}, with Boltzmann weights satisfying the star-star relation, by using the property that the latter relation is equivalent to a particular Seiberg duality for the $4\!-\!d$ $\mathcal{N}=1$ lens index for a class of $SU(N)$ quiver gauge theories.  This star-star relation is rather general and contains its variant for the master solution \cite{Bazhanov:2010kz} and its multi-spin generalisation \cite{BS11,BKS} as particular cases.

In Section \ref{sec:str} it is shown that the Boltzmann weights for the model with two-component spins introduced by Yamazaki, also satisfy a star-triangle relation.  A proof for the star-triangle relation is given in Appendix \ref{app:proof}, which also verifies the corresponding two component spin star-star relation, since the former relation implies the latter (but the reverse is not true).  The actual proof given in Appendix \ref{app:proof} is for an identity more general than the star-triangle relation, resulting in a new elliptic hypergeometric summation/integration identity for the lens elliptic gamma function, that contains six complex and six integer variables.  This identity contains Spiridonov's celebrated elliptic beta integral as a particular case.  Two limiting cases of the star-triangle relation are considered, resulting in one new solution to the star-triangle relation with Boltzmann weights given in terms of infinite q-products, and another new solution with Boltzmann weights given in terms of the Euler gamma function, that has recently been obtained by the author \cite{K14}.  Possible relations to existing integrals in the literature are discussed.

\subsection{Solvable square lattice model}
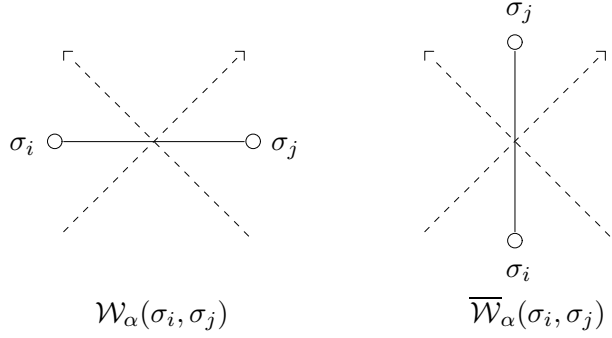
\begin{figure}[hbt]
\begin{center}
\setlength{\unitlength}{1cm}
\begin{picture}(15,5)
\put(3.8,1.5)
{\begin{picture}(6,2)
 \setlength{\unitlength}{1.2cm}
 \path(0,1)(2,1)
 \put(-0.1,1){\circle{0.16}}
 \put(2.1,1){\circle{0.16}}
 \multiput(0,0)(0.11,0.11){18}{\line(1,1){0.05}} \path(2,1.9)(2,2)(1.9,2)
 \multiput(2,0)(-0.11,0.11){18}{\line(1,-1){0.05}} \path(0,1.9)(0,2)(0.1,2)
 \put(-0.6,0.9){$\sigma_i$}
 \put(2.3,0.9){$\sigma_j$}
 \put(0.35,-1.0){$\iW_{\alpha}(\sigma_i,\sigma_j)$}
 \multiput(4,0)(0.11,0.11){18}{\line(1,1){0.05}} \path(6,1.9)(6,2)(5.9,2)
 \multiput(6,0)(-0.11,0.11){18}{\line(1,-1){0.05}} \path(4,1.9)(4,2)(4.1,2)
 \path(5,0)(5,2)
 \put(5,-0.1){\circle{0.16}}
 \put(5,2.1){\circle{0.16}}
 \put(4.9,-0.5){$\sigma_i$}
 \put(4.9,2.4){$\sigma_j$}
 \put(4.5,-1.0){$\overline{\iW}_{\alpha}(\sigma_i,\sigma_j)$}
\end{picture}}
\end{picture}
\caption{Horizontal (left) and vertical edges (right), and their Boltzmann weights.  Here ``rapidity lines'' (dashed arrows) are shown to distinguish the two types of weights, $\iW$ and $\overline{\iW}$.}
\label{2boltzmannweights}
\end{center}
\end{figure}

All models may be considered here on the square lattice made up of of $N$ sites.  Spin variables
\beq
\sigma_j=(x_j,m_j)\,,\quad x_j\in\mathbb{R}\,,\;m_j\in\mathbb{Z}\,,\quad j=1,2,\ldots,N,
\eeq
are assigned to each site of the lattice, where $x_j$ takes real values, and $m_j$ takes integer values.  Two spins interact only if they are connected by an edge of the lattice.  The interactions are represented by the Boltzmann weights $\w_\alpha(\sigma_i,\sigma_j)$, and $\ow_\alpha(\sigma_i,\sigma_j)$, associated to horizontal and vertical edges respectively, where $\sigma_i$ and $\sigma_j$ are the spins located at the end of the edge, as shown in Figure~\ref{2boltzmannweights}.  Here two Boltzmann weights are distinguished by crossing of dashed rapidity lines, a property which allows one to also consider the model on more general ``Z-invariant'' lattices \cite{Bax1}.
The two edge Boltzmann weights, depend on the additive spectral variable $\alpha$, and are related to each other by the crossing symmetry property $\ow_\alpha(\sigma_i,\sigma_j)=\w_{\eta-\alpha}(\sigma_i,\sigma_j)$.  The ``crossing parameter'', $\eta$, is model dependent, and the regime $0<\alpha<\eta$ is a physical regime of the model, where the Boltzmann weights are positive and real-valued.  For all models considered in this paper, the Boltzmann weights are also spin reflection symmetric, such that $\w_\alpha(\sigma_i,\sigma_j)=\w_\alpha(\sigma_j,\sigma_i)$.

To each spin $\sigma_j$ in the lattice one also associates the single-spin weights $\s(\sigma_j)$, which are independent of the spectral variable $\alpha$.  The partition function of the model is then defined as a product of all Boltzmann weights, with a integral (sum) over all internal continuous (discrete) spins, while boundary spins are kept fixed,
\begin{align}
\label{z-main}
{\cal Z}=\sum\int
\prod_{(ij)}\w_{\alpha}(\sigma_i,\sigma_j)\
\prod_{(kl)}\w_{\cpar-\alpha}(\sigma_k,\sigma_l)\ \prod_{n}
\s(\sigma_n)\,d x_n\,.
\end{align}
The first product is taken over all horizontal edges $(ij)$, the second over all vertical edges $(kl)$ and the third product over all internal sites of the lattice.  The goal of statistical mechanics is to evaluate \eqref{z-main} in the thermodynamic limit, when the number of sites in the lattice goes to infinity.  This evaluation is possible if the Boltzmann weights satisfy the star-triangle equation.  For the models given here this relation reads
\begin{align}
\label{msstr}
\begin{array}{l}
\ds\sum_{m_0}\int
 d x_0\,\iS(\sigma_0)\iW_{\eta-\alpha_i}(\sigma_i,\sigma_0)\iW_{\eta-\alpha_j}(\sigma_j,\sigma_0)\iW_{\eta-\alpha_k}(\sigma_0,\sigma_k)\\[.3cm]
\phantom{MMMMMMMMM}\ds=
{\cal
  R}(\alpha_i,\alpha_j,\alpha_k)\,\iW_{\alpha_i}(\sigma_j,\sigma_k)\iW_{\alpha_j}(\sigma_i,\sigma_k)\iW_{\alpha_k}(\sigma_j,\sigma_i)\,,
\end{array}
\end{align}
where the three spectral parameters $\alpha_i$, $\alpha_j$, $\alpha_k$ satisfy the constraint $\alpha_i+\alpha_j+\alpha_k=\eta$ and the factor ${\cal   R}(\alpha_i,\alpha_j,\alpha_k)$ is independent of the spins $\sigma_i,\sigma_j,\sigma_k$.  The integral and sum are evaluated over a given set of continuous and discrete values respectively.  There exists also a second star-triangle relation obtained by exchanging the order of spins appearing in the Boltzmann weights in \eqref{msstr}, however for models with symmetric Boltzmann weights, $\w_\alpha(\sigma_i,\sigma_j)=\w_\alpha(\sigma_j,\sigma_i)$, the two expressions are equivalent.  The star-triangle relation \eqref{msstr} implies that the row-to-row transfer matrices of the lattice model commute \cite{Bax72}.

For all models considered here, the normalisation of the Boltzmann weights is chosen such that the spin independent factor $\mathcal{R}(\alpha_i,\alpha_j,\alpha_k)$ in \eqref{msstr} is equal to one.  This result is based on a factorisation for $\mathcal{R}(\alpha_i,\alpha_j,\alpha_k)$ \cite{Bax02rip}, which holds for all of the above mentioned solutions of the star-triangle relation, particularly for the new solutions appearing in the following section.

For this special normalisation, the Boltzmann weights of the model also satisfy the following boundary conditions
\begin{align}
\label{msbc}
\left.\w_\alpha(\sigma_i,\sigma_j)\right|_{\alpha=0}=1,\quad\left.\w_{\eta-\alpha}(\sigma_i,\sigma_j)\right|_{\alpha\rightarrow0}=\frac{1}{2\s(\sigma_i)}(\delta(x_i\!+\!x_j)\,\delta_{m_i,-m_j}+\delta(x_i\!-\!x_j)\,\delta_{m_i,m_j})\,,
\end{align}
where $\delta(x)$, and $\delta_{m,n}$, are respectively Dirac and Kronecker delta functions.  The exact form of the second boundary condition differs slightly depending on the symmetries of the Boltzmann weights, and explicit expressions for the boundary conditions in each case will be given for the three different models obtained in the next section.

From the boundary conditions \eqref{msbc}, and star-triangle relation \eqref{msstr}, one obtains the following inversion relations
\begin{align}
\label{msinv}
\begin{array}{rcl}
\ds\w_\alpha(\sigma_i,\sigma_j)\w_{-\alpha}(\sigma_i,\sigma_j)&\!\!\!\!=\!\!\!\!&1\,, \\[0.2cm]
\ds\sum_{m_0}\!\int\!\! dx_0\,\s(\sigma_0)\w_{\eta-\alpha}(\sigma_i,\sigma_0)\w_{\eta+\alpha}(\sigma_0,\sigma_j)&\!\!\!\!=\!\!\!\!&\ds\frac{1}{2\s(\sigma_i)}(\delta(x_i\!+\!x_j)\,\delta_{m_i,-m_j}+\delta(x_i\!-\!x_j)\,\delta_{m_i,m_j})\,.
\end{array}
\end{align}
The above relations, \eqref{msstr} and \eqref{msinv}, allow one to show that in the thermodynamic limit as the number of lattice sites goes to infinity $N\to\infty$, the bulk free energy of the model vanishes
\begin{align}
 \lim_{N\to \infty} N^{-1} \log{\cal  Z} = 0\,.\label{fzero}
\end{align}
A derivation of this result requires some extensions \cite{BKS2} of the standard inversion relation method \cite{Str79,Zam79,Bax82inv}.  Here the boundary spins are assumed to be kept finite in the limit $N\to\infty$, and there is an analyticity assumption for the free energy of the model in the physical regime.  The result \eqref{fzero} is purely a consequence of the special choice of normalisation for the Boltzmann weights \cite{BMS07a,BMS07b,Bazhanov:2010kz}.

\section{New discrete and continuous spin solutions to the star-triangle relation}
\label{sec:str}
In this section the Boltzmann weights are defined that give a new solution of the star-triangle relation \eqref{msstr}.\footnote{These Boltzmann weights correspond to the 2-component spin version of Yamazaki's star-star relation, which was obtained by identifying the correspondence of this relation with Seiberg duality of the lens index for a $\mathcal{N}=1$ supersymmetric quiver gauge theory \cite{Yamazaki2013}.}  This star-triangle relation and the corresponding proof given in Appendix \ref{app:proof} are the main result of the paper.

Recall the definition of the spin
\beq
\sigma_j=(x_j,m_j),\quad x_j\in\mathbb{R}\,,\; m_j\in\mathbb{Z}\,.
\eeq
Restrict the continuous real valued component $x_j$, and the discrete integer valued component $m_j$, to take values
\beq
\label{rdef}
0\leq x_j<\pi,\quad m_j=0,1,\ldots,\floor{r/2}\,,
\eeq
for some positive integer parameter $r=1,2,\ldots,$ where $\floor{~}$ is the floor function.  Define also the elliptic nomes $\p,\q$, and crossing parameter $\eta$ as
\beq
\label{nomes}
\p=\EXP^{\pi\ii\sigma},\;\q=\EXP^{\pi\ii\tau},\;\eta=-\pi\ii(\sigma+\tau)/2\,,\quad\im\sigma,\;\im\tau >0\,.
\eeq
Note that a physical regime where Boltzmann weights are real and positive valued can be found for $\p=\q^*$.  Define the elliptic gamma function as \cite{Rui-EGF,BS11}
\beq
\label{egf}
\Phi(z;\p,\q)=\prod_{j,k=0}^\infty\frac{1-\EXP^{2\ii z}\,\p^{2j+1}\,\q^{2k+1}}{1-\EXP^{-2\ii z}\,\p^{2j+1}\,\q^{2k+1}}\,.
\eeq
In terms of the elliptic gamma function, the so-called lens elliptic gamma function is defined as \cite{Yamazaki2013}
\beq
\label{legf}
\begin{array}{rcl}
\ds\Phi_{r,m}(z)\!\!\!\!&=&\!\!\!\!\ds\Phi(z+(r/2-\llbracket m\rrbracket _r)\,\pi\sigma;\p\,\q,\,\p^r)\,\Phi(z-(r/2-\llbracket m\rrbracket _r)\,\pi\tau;\p\,\q,\,\q^r) \\[0.3cm]
\!\!\!\!&=&\!\!\!\!\ds\prod_{j,k=0}^\infty\!\frac{1-\EXP^{2\ii z}\,\p^{-2\llbracket m\rrbracket _r}\,(\p\q)^{2j+1}\,(\p^r)^{2k+2}}{1-\EXP^{-2\ii z}\,\p^{2\llbracket m\rrbracket _r}\,(\p\q)^{2j+1}\,(\p^r)^{2k}}\frac{1-\EXP^{2\ii z}\,\q^{2\llbracket m\rrbracket _r}\,(\p\q)^{2j+1}\,(\q^r)^{2k}}{1-\EXP^{-2\ii z}\,\q^{-2\llbracket m\rrbracket _r}\,(\p\q)^{2j+1}\,(\q^r)^{2k+2}}\,,
\end{array}
\eeq
where $\llbracket  m\rrbracket _r\in\{0,1,\ldots ,r-1\}$ denotes $m \mbox{ modulus } r$.  From \eqref{rdef}, note that when $r=1$, then $m_j=0$, and the lens elliptic gamma function reduces to the usual elliptic gamma-function \eqref{egf}
\beq
\Phi_{1,0}(z)=\Phi(z;\p,\q)\,.
\eeq
The lens elliptic gamma function \eqref{legf}, satisfies the following periodicity and inversion relations
\beq
\Phi_{r,m}(z)=\Phi_{r,m}(z+\pi),\quad\frac{1}{\Phi_{r,m}(z)}=\Phi_{r,-m}(-z)\,.
\eeq
Now define the edge Boltzmann weight as
\beq
\label{ebw}
\w_\alpha(\sigma_i,\sigma_j)=\ds\frac{\EXP^{-2\alpha\,(\,\llbracket m_i-m_j\rrbracket _\pm+\llbracket m_i+m_j\rrbracket _\pm\,)/r}}{\kappa(\alpha)}\frac{\Phi_{r,m_i-m_j}(x_i-x_j+\ii\alpha)\,\Phi_{r,m_i+m_j}(x_i+x_j+\ii\alpha)}{\Phi_{r,m_i-m_j}(x_i-x_j-\ii\alpha)\,\Phi_{r,m_i+m_j}(x_i+x_j-\ii\alpha)}\,,
\eeq
where $\llbracket m\rrbracket_\pm:=\llbracket m\rrbracket_r\llbracket -m\rrbracket_r$.  The spectral parameter $\alpha$ is taken to lie in the domain $0<\alpha<\eta$, where $\eta$ is real. For $\p=\q^*$, this is a physical regime of the model, where the Boltzmann weights \eqref{ebw} take real, positive values.

The normalisation factor $\kappa(\alpha)$ is given by
\beq
\label{ebwnorm}
\kappa(\alpha)=\exp\left\{\sum_{n\neq0}\frac{\EXP^{4\alpha n}((\p\q)^{rn}-(\p\q)^{-rn})}{n((\p\q)^{2n}-(\p\q)^{-2n})(\p^{rn}-\p^{-rn})(\q^{rn}-\q^{-rn})}\right\}\,,
\eeq
and satisfies the pair of functional equations
\beq
\label{functrels}
\frac{\kappa(\eta-\alpha)}{\kappa(\alpha)}=\Phi_{r,0}(\ii(\eta-2\alpha)),\quad\kappa(\alpha)\kappa(-\alpha)=1\,.
\eeq
For $r=1$ this reduces to the normalisation of the Boltzmann weights for the BS master solution \cite{Bazhanov:2010kz}.

Note that the functional equations \eqref{functrels} arise when solving for the free energy of the model, $\lim_{N\rightarrow\infty}N^{-1}\log\cal{Z}$, using the inversion relation method \cite{Str79,Zam79,Bax82inv}.  The solution of these functional equations with appropriate analyticity properties, \eqref{ebwnorm}, is included in the normalisation of the Boltzmann weights \eqref{ebw}, and the result \eqref{fzero} follows \cite{BMS07a,BMS07b,Bazhanov:2010kz,BKS2}.

Next define the single-spin Boltzmann weight as\footnote{This differs from Yamazaki's single-spin weight ($\mathbb{S}^v(s)$ in his notation \cite{Yamazaki2013}) by dropping a constant singular factor that appears to be incorrect, at least from the point of view of the statistical mechanical model (in his notation this factor is $(\Gamma_{r,0}(1;p,q))^{-(N-1)}$).}
\beq
\label{ebws}
\begin{array}{rcl}
\ds\s(\sigma_i)\!\!\!&=&\!\!\!\ds\frac{\varepsilon_i}{\pi}\,(\p^{2r};\p^{2r})_\infty(\q^{2r};\q^{2r})_\infty\,\EXP^{2\eta\llbracket 2m_i\rrbracket _\pm/r}\,\Phi_{r,-2m_i}(-2x_i-\ii\eta)\,\Phi_{r,2m_i}(2x_i-\ii\eta)\,, \\[0.5cm]
\!\!\!&=&\!\!\!\ds\frac{\varepsilon_i}{\pi}\,\EXP^{2\eta\llbracket 2m_i\rrbracket _\pm/r}\,{\vartheta}_4(2x_i+(r/2-\llbracket 2m_i\rrbracket _r)\pi\sigma\,|\,\p^r)\,{\vartheta}_4(2x_i-(r/2-\llbracket 2m_i\rrbracket _r)\pi\tau\,|\,\q^r)\,,
\end{array}
\eeq
where
\beq
\label{epsdef}
\varepsilon_i=\left\{\begin{array}{ll}\frac{1}{2}&\ds\quad m_i=0 \mbox{ or } \llbracket r-m_i\rrbracket _r\,, \\[0.3cm] 1&\quad\mbox{otherwise}\,,\end{array}\right.
\eeq
$\vartheta_4$ is a Jacobi theta function
\beq
\vartheta_4(z\,|\,\p)=(\p^2;\p^2)_\infty\prod_{n=1}^\infty\left(1-\EXP^{2\ii z}\p^{2n-1}\right)\left(1-\EXP^{-2\ii z}\p^{2n-1}\right)\,,
\eeq
and $(x;\q)_\infty=\prod_{j=0}^\infty\,(1-x\,\q^j)$ is the $\q$-Pochhammer symbol.

The Boltzmann weights \eqref{ebw} are reflection symmetric
\beq
\label{spinrefl}
\w_\alpha(\sigma_i,\sigma_j)=\w_\alpha(\sigma_j,\sigma_i)\,.
\eeq
The Boltzmann weights have an obvious $\pi$-periodic symmetry in the continuous spin variable, and they also are invariant under the spin transformation $x_i\rightarrow-x_i$, $m_i\rightarrow r-m_i$,
Accordingly the discrete spins are restricted to values $0,1,\ldots,\floor{r/2}$, and the $\varepsilon_i$ factor was introduced in \eqref{epsdef} to account for this.

The Boltzmann weights \eqref{ebw} satisfy the following boundary conditions analogous to \eqref{msbc}
\begin{align}
\begin{array}{rcl}
\ds\left.\w_\alpha(\sigma_i,\sigma_j)\right|_{\alpha=0}&\!\!\!\!=\!\!\!\!&\ds1,\\[0.3cm]
\ds\left.\w_{\eta-\alpha}(\sigma_i,\sigma_j)\right|_{\alpha\rightarrow0}&\!\!\!\!=\!\!\!\!&\ds\frac{\varepsilon_{i}}{\s(\sigma_i)}(\delta(\sin(x_i\!+\!x_j))\,\delta_{\llbracket m_i+m_j\rrbracket_r,0}+\delta(\sin(x_i\!-\!x_j))\,\delta_{\llbracket m_i-m_j\rrbracket_r,0})\,,
\end{array}
\end{align}
where $\s(\sigma_i)\neq0$.  The $r=1$ case of these relations were previously obtained for the BS master solution \cite{Bazhanov:2010kz}, and in connection with biorthogonality of elliptic hypergeometric functions \cite{Sp2008}.

The Boltzmann weights \eqref{ebw}, and \eqref{ebws}, satisfy the star-triangle relation\footnote{It follows from \eqref{functrels} that a factor $\mathcal{R}(\alpha_i,\alpha_j,\alpha_k)=\Phi_{r,0}(\ii(\eta-2\alpha_i))\Phi_{r,0}(\ii(\eta-2\alpha_j))\Phi_{r,0}(\ii(\eta-2\alpha_k))\,,$ would appear on the right hand side of the star-triangle relation \eqref{str}, if the Boltzmann weights \eqref{ebw} weren't normalised by $\kappa(\alpha)$.}
\beq
\label{str}
\begin{array}{l}
\ds\sum_{m_0=0}^{\floor{r/2}}\,\int^\pi_0\! dx_0\,\iS(\sigma_0)\iW_{\eta-\alpha_i}(\sigma_i,\sigma_0)\iW_{\eta-\alpha_j}(\sigma_j,\sigma_0)\iW_{\eta-\alpha_k}(\sigma_k,\sigma_0)=\iW_{\alpha_i}(\sigma_j,\sigma_k)\iW_{\alpha_j}(\sigma_i,\sigma_k)\iW_{\alpha_k}(\sigma_j,\sigma_i)\,,
\end{array}
\eeq
with the spectral parameters satisfying $\eta=\alpha_i+\alpha_j+\alpha_k$.  For $r=1$ this reduces to the master solution of the star-triangle relation \cite{Bazhanov:2010kz}.

The star-triangle relation \eqref{str} is a particular case of a new elliptic hypergeometric summation/integration identity given in Appendix \ref{app:proof}.

\subsection{Limit: $r\rightarrow\infty$}
\label{sec:rinf}

The $r\rightarrow\infty$ limit of \eqref{str} is formally fairly straightforward due to the simple asymptotics of the lens elliptic gamma function.  Consider the same elliptic nomes $\p,\q$ from the previous section in \eqref{nomes}.  Define the function $Q$ as the $r\rightarrow\infty$ limit of the lens elliptic gamma function \eqref{legf}
\beq
\label{qdef}
Q(z,n)=\lim_{r\rightarrow\infty}\Phi_{r,n}(z)=\left\{
\begin{array}{lr}
\ds\prod_{j=0}^\infty\,\frac{1-\EXP^{2\ii z}\,\p^{-2n}\,(\p\q)^{2j+1}}{1-\EXP^{-2\ii z}\,\q^{-2n}\,(\p\q)^{2j+1}}& n<0\,, \\
\ds\prod_{j=0}^\infty\,\frac{1-\EXP^{2\ii z}\,\q^{2n}\,(\p\q)^{2j+1}}{1-\EXP^{-2\ii z}\,\p^{2n}\,(\p\q)^{2j+1}}& n\geq0\,.
\end{array}\right.
\eeq
This function satisfies the following inversion relation
\beq
Q(z,n)=\frac{1}{Q(-z,-n)}\,.
\eeq
From this function one defines the edge Boltzmann weights
\beq
\label{rinfwts}
\w_\alpha(\sigma_i,\sigma_j)=\frac{\EXP^{-2\alpha |m_i-m_j|-2\alpha|m_i+m_j|}}{\kappa(\alpha)}\,\frac{Q(x_i-x_j+\ii\alpha,m_i-m_j)\,Q(x_i+x_j+\ii\alpha,m_i+m_j)}{Q(x_i-x_j-\ii\alpha,m_i-m_j)\,Q(x_i+x_j-\ii\alpha,m_i+m_j)}\,,
\eeq
with the normalisation
\beq
\label{rinfnorm}
\kappa(\alpha)=\exp\left\{-\sum_{n\neq0}\frac{\EXP^{4\alpha n}}{n((\p\q)^{2n}-(\p\q)^{-2n})}\right\}\,.
\eeq
The spectral variable is restricted to the region $0<\alpha<\eta$, with $\eta$ defined in \eqref{nomes}.  The normalisation factor $\kappa$ satisfies the following functional equations
\beq
\frac{\kappa(\eta-\alpha)}{\kappa(\alpha)}=Q(\ii(\eta-2\alpha),0),\quad\kappa(\alpha)\kappa(-\alpha)=1\,,
\eeq
which are required for \eqref{fzero} to hold.

Define also the single-spin Boltzmann weight as
\beq
\label{rinfswt}
\s(\sigma_j)=\frac{1}{2\pi}\,\EXP^{4\eta |m_j|}\,Q(2x_j-\ii\eta,2m_j)\,Q(-2x_j-\ii\eta,-2m_j)\,.
\eeq
The continuous spins $x_j$ and discrete spins $m_j$ now take values
\beq
0\leq x_j<\pi,\quad m_j\in\mathbb{Z}\,.
\eeq
The Boltzmann weights \eqref{rinfwts} satisfy spin reflection symmetry
\beq
\w_\alpha(\sigma_i,\sigma_j)=\w_\alpha(\sigma_j,\sigma_i)\,,
\eeq
and are $\pi$-periodic in the continuous spin $x_j$.  These Boltzmann weights are real and positive for $\p=\q^*$, and $0<\alpha<\eta$.

The Boltzmann weights \eqref{rinfwts} satisfy the following boundary conditions
\begin{align}
\left.\w_\alpha(\sigma_i,\sigma_j)\right|_{\alpha=0}=1,\quad\left.\w_{\eta-\alpha}(\sigma_i,\sigma_j)\right|_{\alpha\rightarrow0}=\frac{1}{2\s(\sigma_i)}(\delta(\sin(x_i\!+\!x_j))\,\delta_{m_i,-m_j}+\delta(\sin(x_i\!-\!x_j))\,\delta_{m_i,m_j})\,,
\end{align}
where $\s(\sigma_i)\neq0$.

The Boltzmann weights \eqref{rinfwts}, and \eqref{rinfswt}, satisfy the star-triangle relation
\beq
\label{rinfstr}
\begin{array}{r}
\ds\sum_{m_0\in\mathbb{Z}}\,\int^\pi_0\! dx_0\,\iS(\sigma_0)\iW_{\eta-\alpha_i}(\sigma_i,\sigma_0)\iW_{\eta-\alpha_j}(\sigma_j,\sigma_0)\iW_{\eta-\alpha_k}(\sigma_k,\sigma_0)=\iW_{\alpha_i}(\sigma_j,\sigma_k)\iW_{\alpha_j}(\sigma_i,\sigma_k)\iW_{\alpha_k}(\sigma_j,\sigma_i)\,,
\end{array}
\eeq
with $\eta=\alpha_i+\alpha_j+\alpha_k$.

Note that a similar but different identity, involving an integral and sum over continuous and discrete variables respectively, was recently obtained by Gahramanov and Rosengren in the form of a pentagon identity from $3\!-\!d$ $\mathcal{N}=2$ supersymmetric gauge theories \cite{GR13}.  From this point of view, it would also be interesting to find an interpretation of equation \eqref{rinfstr}, if any, in terms of a new duality for the $r\rightarrow\infty$ reduction of the $S_1\times S_3/\mathbb{Z}_r$ superconformal indices \cite{BNY13,Y14}.

\subsection{Gamma function limit}

From the star-triangle relation \eqref{rinfstr}, one can take a further limit to obtain another new solution of the star-triangle relation, with Boltzmann weights given in terms of the Euler gamma function.  The latter star-triangle relation was recently obtained by the author \cite{K14}.

Consider the following limit of the elliptic nomes 
\beq
\p=\EXP^{-\hbar},\;\;\q=\EXP^{-\hbar},\;\;\eta=\hbar,\quad\hbar\rightarrow0 \,,
\eeq
and the following scaling limit of the continuous spins $x_j$ and spectral parameters $\alpha$ from Section \ref{sec:rinf}
\beq
x_j\rightarrow\hbar x_j,\;\;\alpha\rightarrow\hbar\alpha\,,\quad\hbar\rightarrow0\,.
\eeq
Under the rescaling of the spins $\sigma_j=(x_j\hbar,m_j)$, the asymptotics of \eqref{qdef}, in the previous section as $\hbar\rightarrow0$, are given in terms of the Euler gamma function $\Gamma(z)$ by
\beq
Q(\sigma_j)\simeq(4\hbar)^{\ii x_j}\,\frac{\Gamma(\frac{1+|m_j|+\ii x_j}{2})}{\Gamma(\frac{1+|m_j|-\ii x_j}{2})}\,.
\eeq
Then as $\hbar\rightarrow0$, the asymptotics of the Boltzmann weights \eqref{rinfwts}, \eqref{rinfswt}, and normalisation \eqref{rinfnorm}, are given by
\beq
\s(\sigma_j)\simeq\frac{1}{2\pi}(4\hbar)^{2}(x_j^2+m_j^2),\quad\kappa(\alpha\hbar)\simeq(8\hbar)^{-\alpha}\,\frac{\Gamma(\frac{1-\alpha}{2})}{\Gamma(\frac{1+\alpha}{2})}\,,
\eeq
and\footnote{The following compact notation for products of the gamma function, $\Gamma(x\pm y)=\Gamma(x+y)\Gamma(x-y)$, is now used for convenience.}
\beq
\kappa(\alpha\hbar)\,\w_{\alpha\hbar}(\sigma_i,\sigma_j)\simeq(4\hbar)^{-4\alpha}\,\frac{\Gamma(\frac{1-\alpha+(m_i-m_j)\pm\ii(x_i-x_j)}{2})\,\Gamma(\frac{1-\alpha+(m_i+m_j)\pm\ii(x_i+x_j)}{2})}{\Gamma(\frac{1+\alpha+(m_i-m_j)\pm\ii(x_i-x_j)}{2})\,\Gamma(\frac{1+\alpha+(m_i+m_j)\pm\ii(x_i+x_j)}{2})}\,.
\eeq
In the above limit the star-triangle relation \eqref{rinfstr}, formally reduces to the following star-triangle relation
\beq
\label{strmsg}
\begin{array}{r}
\ds\sum_{m_0\in\mathbb{Z}}\,\int^{\infty}_{-\infty}\! dx_0\,
\,\iS(\sigma_0)\iW_{\eta-\alpha_i}(\sigma_i,\sigma_0)\iW_{\eta-\alpha_j}(\sigma_j,\sigma_0)\iW_{\eta-\alpha_k}(\sigma_k,\sigma_0)=\iW_{\alpha_i}(\sigma_j,\sigma_k)\iW_{\alpha_j}(\sigma_i,\sigma_k)\iW_{\alpha_k}(\sigma_j,\sigma_i)\,,
\end{array}
\eeq
where $\eta=1=\alpha_i+\alpha_j+\alpha_k$, and\footnote{A typo appeared in the Boltzmann weight $\iS_(\sigma_j)$ in a previous paper \cite{K14}, it has been corrected here by the addition of the factor of $\frac{1}{2}$.} 
\beq
\label{2spinwts}
\iS(\sigma_j)=\ds\frac{1}{4\pi}(x_j^2+m_j^2)\;,\quad\iW_{\alpha}(\sigma_i,\sigma_j)
=\ds\frac{\Gamma(\frac{1+\alpha}{2})}{\Gamma(\frac{1-\alpha}{2})}\,
\frac{\Gamma(\frac{1-\alpha-(m_i+m_j)\pm\ii(x_i+x_j)}{2})\,\Gamma(\frac{1-\alpha-(m_i-m_j)\pm\ii(x_i-x_j)}{2})}
{\Gamma(\frac{1+\alpha-(m_i+m_j)\pm\ii(x_i+x_j)}{2})\,\Gamma(\frac{1+\alpha-(m_i-m_j)\pm\ii(x_i-x_j)}{2})}\;.
\eeq
The spins of the model now take their values $x_i\in\mathbb{R}$, $m_i\in\mathbb{Z}\,,$ and the spectral parameter is restricted to $0<\alpha<\eta$, which is a physical regime of the model.  These Boltzmann weights also obey the spin reflection identity \eqref{spinrefl}, however are no longer $\pi$-periodic in the spin.

The Boltzmann weights \eqref{2spinwts} also satisfy the following boundary conditions
\begin{align}
\left.\w_\alpha(\sigma_i,\sigma_j)\right|_{\alpha=0}=1,\quad\left.\w_{\eta-\alpha}(\sigma_i,\sigma_j)\right|_{\alpha\rightarrow0}=\frac{1}{2\s(\sigma_i)}(\delta(x_i\!+\!x_j)\,\delta_{m_i,-m_j}+\delta(x_i\!-\!x_j)\,\delta_{m_i,m_j})\,,
\end{align}
where $\s(\sigma_i)\neq0$.  For additional details on this star-triangle relation, the reader is referred to the previous publication \cite{K14}.

Note that different solutions to the star-triangle relation with Boltzmann weights given in terms of the Euler gamma function, were previously found in relation to the chiral Potts model \cite{AuYangPerk:ninfcp}.  The Boltzmann weights for these star-triangle relations are for models that contain only, either continuous, or discrete valued spins.  It would be interesting to determine if there exists some relation between these solutions, and the star-triangle relation given in \eqref{strmsg}.

As was previously remarked \cite{K14}, the appearance of discrete and continuous valued spins here resemble the elliptic model obtained by Yamazaki from quiver gauge theory \cite{Yamazaki2013}.  It has been shown here now, how these models are connected through the star-triangle relation \eqref{str}, the latter relation implying the star-star relation for the two-component spin case.

The alternate method \cite{K14} to obtain \eqref{strmsg} was to use a scaling limit of the hyperbolic beta integral solution of the star-triangle relation \cite{Spiridonov-statmech}, that resulted in the star-triangle relation given by \eqref{strmsg}.\footnote{Bazhanov, Mangazeev, and Sergeev originally used this limiting procedure to obtain a related solution to the star-triangle relation, with Boltzmann weights depending only on the differences of spins.\cite{BMS07b}}  The asymptotics of the hyperbolic beta integral in the strong coupling regime, are such that sharp delta function shaped peaks appear when the real valued spins take integer values.  These asymptotics are manifest in the strong coupling limit as additional discrete integer spin variables, as appearing in \eqref{strmsg}.  One might then ask whether the elliptic variant \eqref{str} of this star-triangle relation, arises in the strong coupling limit of some as yet unknown star-triangle relation.  Such a relation should also then have implications for supersymmetric gauge theories, as well as providing a case of an interesting new summation/integration identity, perhaps in terms of more general special functions.

\section{Conclusion}
A new solution to the star-triangle relation was given in \eqref{str}, for an Ising type model whose spins contain integer and real valued components.  The Boltzmann weights of this model are obtained from Yamazaki's Gauge/YBE correspondence and the related solution to the star-star relation \cite{Yamazaki2013}.  The star-triangle relation \eqref{str} implies the two component spin case of the star-star relation given by Yamazaki.  In Appendix \ref{app:proof} a proof is presented of a new elliptic hypergeometric summation/integration identity which contains the star-triangle relation \eqref{str} as a particular case.  The new identity contains six integer variables, in addition to six complex variables, and contains Spiridonov's elliptic beta integral \cite{Spiridonov-beta} as a particular case.

Two further solutions of the star-triangle relation \eqref{rinfstr}, \eqref{strmsg} were given that arise as limiting cases of \eqref{str}.  The Boltzmann weights for these star-triangle relations similarly describe Ising type models with integer and real valued spin components. The star-triangle relation \eqref{rinfstr} appears to be new, while the star-triangle relation \eqref{strmsg} was previously obtained by the author, using a different limiting case.  In each case the Boltzmann weights are normalised such that \eqref{fzero} holds for the corresponding lattice model.

It would be interesting to determine the exact role of the three star-triangle relations \eqref{str}, \eqref{rinfstr}, and \eqref{strmsg}, in the gauge theory setting.  The Gauge/YBE duality described by Yamazaki, implies that the star-triangle relation \eqref{str} should correspond to Seiberg duality of the $4\!-\!d$ $\mathcal{N}=1$ $S_1\times S_3/\mathbb{Z}_r$ index for $SU(2)$ quiver gauge theory.  While the star-triangle relation \eqref{rinfstr} appears to be related to the pentagon identity recently obtained by Gahramanov and Rosengren \cite{GR13}, and is thus expected to correspond to a duality of indices in $3\!-\!d$ $\mathcal{N}=2$ supersymmetric gauge theory \cite{BNY13,Y14}.  

It would also be of interest to mathematically prove star-star relations for multi-component spins given by Yamazaki, that correspond to multivariate generalisations of elliptic hypergeometric integral identities of the type in \eqref{str3}.  It may be possible to do this by adapting proofs given by Rains for the continuous variable cases \cite{Rains-transformations}, and would likely result in new elliptic hypergeometric identities involving an integral and sum over continuous and discrete variables respectively.  

\section*{Acknowledgments}
I thank Vladimir Bazhanov for suggesting the problem, useful advice, and reading the manuscript.  I thank Alexander Bobenko and Yuri Suris for fruitful discussions and their hospitality during my stay at the Technische Universit\"{a}t Berlin.  I also thank Hjalmar Rosengren for his interest to this work and interesting correspondence.  Note that a star-triangle relation closely related to \eqref{rinfstr} was indpendently found \cite{GahSpi}, which became known to the author after these results were completed.  The author is supported by the DFG Collaborative Research Center TRR 109, ``Discretization in Geometry and Dynamics''.

\app{Proof of \eqref{str}}
\label{app:proof}

In this section a proof of \eqref{str} is given.  This is based on and follows closely Spiridonov's proofs of the elliptic beta integrals \cite{Spiridonov-proofs,Spiridonov-essays}.\footnote{See also Wilf and Zeilberger's work \cite{WilfZeil}.}  One major difference is that rather than considering the integral over a closed contour encircling the origin, the integral is considered over the interval $[0,2\pi]$ (the difference between the contours is a simple change of variables).  This is done for convenience, primarily to avoid calculations involving roots of complex numbers.

Recall the definition of the elliptic nomes
\beq
\p=\EXP^{\ii\pi\sigma}\,,\quad\q=\EXP^{\ii\pi\tau}\,,\quad\EXP^{-2\eta}=\p\q\,,\quad\im(\sigma),\;\im(\tau) >0\,,
\eeq
and now define
\beq
\zeta=\ii\pi(1+\tau/2-\sigma/2)\,,
\eeq
and the following function
\beq
\varphi(z,m)=\left(-2\eta-2\ii z+2\zeta(\llbracket m\rrbracket -\llbracket-m\rrbracket)/3\right)\llbracket m\rrbracket _\pm/(4r)\,,
\eeq
with $z\in\mathbb{C}$, $m\in\mathbb{Z}$, $r$ defined in \eqref{rdef}, and $\llbracket m\rrbracket\in\{0,1,\ldots,r-1\}$ denotes $m$ modulus $r$.\footnote{This is equivalent to the definition of $\llbracket m\rrbracket_r$ from Section \ref{sec:str}, with the $r$ subscript now dropped}  Define $\Gamma$ to be the lens elliptic gamma function \cite{Yamazaki2013} in the following form\footnote{This is related to $\Phi(z,m)$ of \eqref{legf} by a change of variables $\EXP^{-\varphi(z,m)}\Gamma(-2z+\ii\eta,m)$, and squaring the elliptic nomes $\p\rightarrow\p^2$, $\q\rightarrow\q^2$.}
\beq
\label{legf2}
\Gamma(z,m)=\EXP^{\varphi(z,m)}\prod_{j,k=0}^\infty\frac{1-\EXP^{-\ii z}\p^{-\llbracket m\rrbracket }(\p\q)^{j+1}\p^{r(k+1)}}{1-\EXP^{\ii z}\p^{\llbracket m\rrbracket }(\p\q)^j\p^{rk}}\frac{1-\EXP^{-\ii z}\q^{-r+\llbracket m\rrbracket }(\p\q)^{j+1}\q^{r(k+1)}}{1-\EXP^{\ii z}\q^{r-\llbracket m\rrbracket }(\p\q)^j\q^{rk}}\,.
\eeq
It is useful to introduce the following compact notation for products of this function
\beq
\Gamma(x \pm z,m\pm n):=\Gamma(x+z,m+n)\,\Gamma(x-z,m-n),\quad \Gamma(x\pm 2z,m\pm 2n):=\Gamma(x+2z,m+2n)\,\Gamma(x-2z,m-2n)\,.
\eeq
The poles of the lens elliptic gamma function \eqref{legf2} are located at the points
\beq
z=-\pi\sigma\left(rj+\llbracket m\rrbracket\right)-2\ii\eta k\,,\,-\pi\tau\left(r(j+1)-\llbracket m\rrbracket\right)-2\ii\eta k\,,
\eeq
and its zeros are located at the points
\beq
z=\pi\sigma\left(r(j+1)-\llbracket m\rrbracket\right)+2\ii\eta(k+1)\,,\,\pi\tau\left(rj+\llbracket m\rrbracket\right)+2\ii\eta(k+1)\,,
\eeq
for $j,k=0,1,\ldots$.

The lens elliptic gamma function \eqref{legf2} obeys the following useful identities
\beq
\Gamma(z,m)=\frac{1}{\Gamma(2\ii\eta-z,-m)}\,,
\eeq
and
\beq
\Gamma(z+n\pi\sigma,m-n)=\left(\prod_{j=0}^{n-1}\theta(z+j\pi\sigma,m-j\,|\,\tau)\right)\,\Gamma(z,m)\,,\quad n=0,1,\ldots\,,
\eeq
where the theta function is defined as
\beq
\label{thtdef}
\theta(z,m\,|\,\tau)=\EXP^{\phi(z,m)}(\EXP^{\ii z}\q^{\llbracket-m\rrbracket};\q^r)_\infty(\EXP^{-\ii z}\q^{r-\llbracket-m\rrbracket};\q^r)_\infty\,,
\eeq
and
\beq
\phi(z,m)=\frac{\varphi(z+\pi\sigma,m-1)}{\varphi(z,m)}=\left(\zeta(r-1)(r+1)/3-\pi\ii(\tau+2)\llbracket m\rrbracket_\pm-\ii(z+\pi)(r-1-2\llbracket-m\rrbracket)\right)/(2r)\,.
\eeq
The theta function \eqref{thtdef} obeys the following useful identities
\beq
\theta(-z,-m\,|\,\tau)=-\EXP^{\ii(2\pi\llbracket m\rrbracket-z)/r}\theta(z,m\,|\,\tau)\,,\quad\theta(z+nr\pi\tau,m\,|\,\tau)=\EXP^{\ii n(\pi-z-\pi\tau(nr-1)/2)}\theta(z,m\,|\,\tau)\,,
\eeq
where $n\in\mathbb{Z}$.

A more general identity than the star-triangle relation \eqref{str}, is the following summation/integration identity\footnote{Spiridonov's elliptic beta integral \cite{Spiridonov-beta} corresponds to the $r=1$ case of this integral.}
\beq
\label{str3}
\ds(\q^r;\q^r)_\infty(\p^r;\p^r)_\infty\,\sum_{y=0}^{r-1}\int_0^{2\pi}\,\frac{dz}{4\pi}\,\frac{\prod_{i=1}^6\Gamma(t_i\pm z,u_i\pm y)}{\Gamma(\pm 2z,\pm 2y)}\ds=\!\!\!\!\prod_{1\leq i<j\leq6}\Gamma(t_i+t_j,u_i+u_j)\,,
\eeq
where
\beq
\p,\q,t_i\in\mathbb{C},\quad u_i\in\mathbb{Z},\quad |\p|,|\q|<1,\quad\im(t_i)>0\,,\quad i=1,\ldots,6\,,
\eeq
and the variables are restricted to satisfy
\beq
\sum_{i=1}^6t_i= 2\ii\eta,\quad\sum_{i=1}^6u_i=0\,.
\eeq
The star-triangle relation \eqref{str} is related to the identity \eqref{str3} by the change of variables
\beq
\label{cov}
\begin{array}{rclrclrcl}
\ds t_1&\!\!=\!\!&\ds{ x_1+\ii \alpha_1}, & t_3&\!\!=\!\!&\ds{ x_3+\ii\alpha_3}, & t_5&\!\!=\!\!&\ds{x_2-\ii(\alpha_1 +\alpha_3 -\eta)}\,, \\
\ds t_2&\!\!=\!\!&\ds{-x_1 +\ii\alpha_1}, & t_4&\!\!=\!\!&\ds{-x_3+\ii\alpha_3}, & t_6&\!\!=\!\!&\ds{-x_2-\ii(\alpha_1+\alpha_3-\eta)}\,,
\end{array}
\eeq
and
\beq
\label{cov}
\begin{array}{rclrclrcl}
\ds u_1&\!\!=\!\!&\ds m_1,& u_3&\!\!=\!\!&\ds m_3, & u_5&\!\!=\!\!& \ds m_2\,, \\
\ds u_2&\!\!=\!\!&\ds -m_1, & u_4&\!\!=\!\!& \ds -m_3, & u_6&\!\!=\!\!& \ds -m_2\,.
\end{array}
\eeq

The identity \eqref{str3} is what is to be proven. This identity can be re-written in the equivalent form
\beq
\label{str2}
I(t_1,\ldots,t_5,u_1,\ldots,u_5)=\sum_{y=0}^{r-1}\,\int_0^{2\pi}\,\rho(z,y,t_1,\ldots,t_5,u_1,\ldots,u_5)\,dz=\frac{4\pi}{(\q^r;\q^r)_\infty(\p^r;\p^r)_\infty}\,,
\eeq
where
\beq
\rho(z,y,t_1,\ldots,t_5,u_1,\ldots,u_5)=\frac{\prod_{i=1}^5\Gamma(t_i\pm z,u_i\pm y)\,\Gamma(A-t_i,U-u_i)}{\Gamma(\pm 2z,\pm 2y)\,\Gamma(A\pm z,U\pm y)\,\prod_{1\leq i<j\leq5}\,\Gamma(t_i+t_j,u_i+u_j)}\,,
\eeq
and 
\beq
A=\sum_{i=1}^5t_i,\quad U=\sum_{i=1}^5u_i\,.
\eeq
The integral \eqref{str3} is then recovered by setting $t_6=2\ii\eta-A$, and $u_6=-U$.

The integrand $\rho$ is $2\pi$-periodic in $z$
\beq
\rho(z+2\pi k,y,t_1,\ldots,t_5,u_1,\ldots,u_5)=\rho(z,y,t_1,\ldots,t_5,u_1,\ldots,u_5),\,
\eeq
for $k\in\mathbb{Z}$.

For $|\im(A)|<|\im(2\ii\eta)|$, the integrand $\rho$ has the following poles lying in the upper half plane
\beq
\label{rhopoles1}
\begin{array}{c}
\ds\left\{t_i+\pi\sigma\left(rj+\llbracket u_i-y\rrbracket\right)+2\ii\eta k+2\pi n,t_i+\pi\tau\left(r(1+j)-\llbracket u_i-y\rrbracket\right)+2\ii\eta k+2\pi n\right\}\,,\\[0.3cm]
\ds\left\{-A+\pi\sigma\left(r(j+1)-\llbracket U+y\rrbracket\right)+2\ii\eta(k+1)+2\pi n,-A+\pi\tau\left(rj+\llbracket U+y\rrbracket\right)+2\ii\eta(k+1)+2\pi n\right\}\,,
\end{array}
\eeq
and the following poles lying in the lower half plane
\beq
\label{rhopoles2}
\begin{array}{c}
\ds\left\{-t_i-\pi\sigma\left(rj+\llbracket u_i+y\rrbracket\right)-2\ii\eta k+2\pi n,-t_i-\pi\tau\left(r(1+j)-\llbracket u_i+y\rrbracket\right)-2\ii\eta k+2\pi n\right\}\,,\\[0.3cm]
\ds\left\{A-\pi\sigma\left(r(j+1)-\llbracket U-y\rrbracket\right)-2\ii\eta(k+1)+2\pi n,A-\pi\tau\left(rj+\llbracket U-y\rrbracket\right)-2\ii\eta(k+1)+2\pi n\right\}\,,
\end{array}
\eeq
for $i=1,2,\ldots,5$, $n\in\mathbb{Z}$, and $j,k=0,1,\ldots$.  By analyticity and periodicity one may also consider the integral \eqref{str2} (or \eqref{str3}), over any contour between the endpoints $z=0,2\pi$ in the strip $0<\re(z)<2\pi$, such that the contour separates the points in the two sets of poles \eqref{rhopoles1} and \eqref{rhopoles2} that lie in this strip, then the $t_i$ can be chosen to be any complex numbers as long as such a contour exists. 

The idea of the proof is to use a difference equation for $\rho$ to show that $I(t_1,\ldots,t_5,u_1,\ldots,u_5)$ is independent of $t_i$ and $u_i$, $i=1,\ldots,5$, and thus only depends on $\p$ and $\q$.  Then one can evaluate $I(t_1,\ldots,t_5,u_1,\ldots,u_5)$ using residues at a special value of $t_i$ and $u_i$, to give \eqref{str2}.

The first step is to establish the following relation
\beq
\label{Ifunct}
I(t_1+\pi\sigma r,t_2,\ldots,t_5,u_1,\ldots,u_5)=I(t_1,\ldots,t_5,u_1,\ldots,u_5)\,.
\eeq

To establish this, observe that $\rho$ satisfies the following difference equation
\beq
\label{funct}
\begin{array}{l}
\ds\rho(z,y,t_1+\pi\sigma,t_2,\ldots,t_5,u_1-1,u_2,\ldots,u_5)-\rho(z,y,t_1,\ldots,t_5,u_1,\ldots,u_5) \\[0.3cm]
\ds\quad=G(z-\pi\sigma,y+1,t_1,\ldots,t_5,u_1,\ldots,u_5)-G(z,y,t_1,\ldots,t_5,u_1,\ldots,u_5)\,,
\end{array}
\eeq
where $G$ is defined as
\beq
\label{gdef}
\begin{array}{l}
\ds G(z,y,t_1,\ldots,t_5,u_1,\ldots,u_5)=\ds\rho(z,y,t_1,\ldots,t_5,u_1,\ldots,u_5)\,\times\\[0.3cm]
\ds \qquad\qquad\qquad\qquad \EXP^{2\ii\pi\llbracket y-u_1\rrbracket/r}\,\frac{\EXP^{\ii t_1/r}}{\EXP^{\ii z/r}} \,\frac{\prod_{i=1}^5\theta(t_i+z,u_i+y\,|\,\tau)}{\prod_{i=2}^5\theta(t_1+t_i,u_1+u_i\,|\,\tau)}\frac{\theta(t_1+A,u_1+U\,|\,\tau)}{\theta(2z,2y\,|\,\tau)\,\theta(A+z,U+y\,|\,\tau)}\,.
\end{array}
\eeq
To see that \eqref{funct} holds, divide both sides of \eqref{funct} by $\rho(z,y,t_1,\ldots,t_5,u_1,\ldots,u_5)$, and one obtains
\beq
\label{thtfunct}
\begin{array}{l}
\ds\frac{\theta(t_1+z,u_1+y\,|\,\tau)\,\theta(t_1-z,u_1-y\,|\,\tau)}{\theta(A+z,U+y\,|\,\tau)\,\theta(A-z,U-y\,|\,\tau)}\prod_{i=2}^5\frac{\theta(A-t_i,U-u_i\,|\,\tau)}{\theta(t_1+t_i,u_1+u_i\,|\,\tau)}-1 \\[0.6cm]
\ds\qquad\qquad=(-)\frac{\EXP^{\ii t_1/r}\,\theta(t_1+A,u_1+U\,|\,\tau)}{\prod_{i=2}^5\theta(t_1+t_i,u_1+u_i\,|\,\tau)}\left(\frac{\EXP^{-\ii z/r+2\ii\pi\llbracket y-u_1\rrbracket/r}\prod_{i=1}^5\theta(t_i+z,u_i+y\,|\,\tau)}{\theta(2z,2y\,|\,\tau)\,\theta(A+z,U+y\,|\,\tau)}+\right. \\[0.6cm]
\ds\left.\qquad\qquad\qquad\qquad\qquad\qquad\EXP^{\ii z/r}\EXP^{2\ii\pi\left(\llbracket y-u_1+1\rrbracket+\llbracket-2y-1\rrbracket\right)/r}\frac{\prod_{i=1}^5\theta(t_i-z,u_i-y\,|\,\tau)}{\theta(-2z,-2y\,|\,\tau)\,\theta(A-z,U-y\,|\,\tau)}\right)\,.
\end{array}
\eeq
Both sides of this relation are elliptic functions of $z$, sharing the same poles and corresponding residues, then from Liouville's theorem, the difference of both sides is a constant.  The constant may be shown to be zero. To check that \eqref{thtfunct} holds is straightforward, and involves simplifying many expressions in terms of $\llbracket m\rrbracket$.  Some more about the identity \eqref{thtfunct} is explained in Appendix \ref{subsec:todo}.

Now integrate both sides of \eqref{funct}, over $0\leq z\leq2\pi$, to obtain
\beq
\label{functinteg}
\begin{array}{l}
\ds I(t_1+\pi\sigma,t_2,\ldots,t_5,u_1-1,u_2,\ldots,u_5)-I(t_1,\ldots,t_5,u_1,\ldots,u_5) \\[0.3cm]
\ds\qquad\qquad\qquad=\sum_{y=0}^{r-1}\left(\int_{-\ii\pi\im(\sigma)}^{2\pi-\ii\pi\im(\sigma)}-\int_0^{2\pi}\right)dz\,G(z,y,t_1,\ldots,t_5,u_1,\ldots,u_5)\,,
\end{array}
\eeq
where $2\pi$-periodicity of $G$ has been used for the first integral on the right hand side.  This first integral is over a straight line connecting the two points $z=-\ii\pi\im(\sigma)$ and $z=2\pi-\ii\pi\im(\sigma)$.  Setting $\im(t_i)>0$, and $\im(A)<\im(\pi\tau)$, the function G in \eqref{gdef} has poles in the upper half plane at the points
\beq
\label{Gpoles1}
\begin{array}{c}
\ds\left\{t_i+\pi\sigma\left(rj+\llbracket u_i-y\rrbracket\right)+2\ii\eta k+2\pi n,t_i+\pi\tau\left(r(1+j)-\llbracket u_i-y\rrbracket\right)+2\ii\eta k+2\pi n\right\}\,,\\[0.3cm]
\ds\left\{-A+\pi\sigma\left(r(j+1)+\llbracket U+y\rrbracket\right)+2\ii\eta(k+1)+2\pi n,-A+\pi\tau\left(r(j+1)+\llbracket U+y\rrbracket\right)+2\ii\eta(k+1)+2\pi n\right\}\,, \\[0.3cm]
\ds\left\{-A+\pi\tau\left(r(j+1)+\llbracket U+y\rrbracket\right)+2\pi n,-A+\pi\tau\left(\llbracket U+y\rrbracket\right)+2\ii\eta (k+1)+2\pi n\right\}\,,
\end{array}
\eeq
and for $\llbracket U+y\rrbracket\neq0$,
\beq
\label{Gpoles11}
\left\{-A+\pi\tau\llbracket U+y\rrbracket+2\pi n\right\}\,,
\eeq
and the following poles in the lower half plane
\beq
\label{Gpoles2}
\begin{array}{c}
\ds\left\{-t_i-\pi\sigma\left(r(j+1)+\llbracket u_i+y\rrbracket\right)-2\ii\eta(k+1)+2\pi n,-t_i-\pi\tau\left(r(j+1)-\llbracket u_i+y\rrbracket\right)-2\ii\eta(k+1)+2\pi n\right\}\,,\\[0.3cm]
\ds\left\{-t_i-\pi\sigma\left(r(j+1)+\llbracket u_i+y\rrbracket\right)-2\ii\eta k+2\pi n,-t_i-\pi\sigma\left(rj+\llbracket u_i+y\rrbracket\right)-2\ii\eta (k+1)+2\pi n\right\}\,,\\[0.3cm]
\ds\left\{A-\pi\sigma\left(r(j+1)-\llbracket U-y\rrbracket\right)-2\ii\eta(k+1)+2\pi n,A-\pi\tau\left(rj+\llbracket U-y\rrbracket\right)-2\ii\eta(k+1)+2\pi n \right\}\,,
\end{array}
\eeq
and for $\llbracket u_i+y\rrbracket\neq0$,
\beq
\label{Gpoles22}
\left\{-t_i-\pi\sigma\llbracket u_i+y\rrbracket+2\pi n\right\}\,,
\eeq
for $i=1,\ldots,5$, $n\in\mathbb{Z}$, and $j,k=0,1,\ldots$.  Since for $\im(t_i)>0$ and $\im(A)<\im(\pi\tau)$, there are no poles in the strip $0>\im(z)>-\im(\pi\sigma)$, one can shift the contour of integration in \eqref{functinteg} from the line $\im(z)=-\im(\pi\sigma)$, to the real axis on $[0,2\pi]$, to obtain zero on the right hand side, thus
\beq
\label{Ifunct2}
I(t_1+\pi\sigma,t_2,\ldots,t_5,u_1-1,u_2,\ldots,u_5)=I(t_1,\ldots,t_5,u_1,\ldots,u_5)\,.
\eeq
Applying this identity $r$ times (as long as parameters remain in our allowed region), one has now obtained Equation \eqref{Ifunct}:
\beq
I(t_1+\pi\sigma r,t_2,\ldots,t_5,u_1,\ldots,u_5)=I(t_1,\ldots,t_5,u_1,\ldots,u_5)\,.
\eeq
In a similar fashion, by requiring that $\im(A)<\im(\pi\sigma)$, one can establish the following difference relation
\beq
I(t_1+\pi\tau r,t_2,\ldots,t_5,u_1,\ldots,u_5)=I(t_1,\ldots,t_5,u_1,\ldots,u_5)\,.
\eeq

For now set $\re(\sigma),\re(\tau)=0$, $\im(\sigma)>\im(\tau)$, and $r\sigma k\neq r\tau k$ for any $n,k=0,1,\ldots$.  Also let the real parts of $A$, and $t_i$, $i=1,\ldots 5$, be non-zero and differ from each other.  For the integral $I(t_1,\ldots,t_5,u_1,\ldots,u_5)$ in \eqref{str2}, deform the contour of integration from $z\in[0,2\pi]$, such that the poles in \eqref{rhopoles1}, and the following points lie above the contour
\beq
\begin{array}{rcl}
\ds\left\{t_1+\ii\pi x\right.&\!\!\!\!|\!\!\!\!&\ds x\geq\min\left(\im(\sigma)\llbracket u_1-y\rrbracket,\im(\tau)(r-\llbracket u_1-y\rrbracket)\right)\,\wedge \\[0.3cm]
&& \left.\ds x\leq\max\left(\im(\sigma)\llbracket u_1-y\rrbracket,\im(\tau)(r-\llbracket u_1-y\rrbracket)\right)+2\im(\sigma)r\right\}\,,\\[0.3cm]
\ds\left\{-A+2\ii\eta+\ii\pi x\right.&\!\!\!\!|\!\!\!\!&\ds x\geq\min\left(\im(\sigma)(r-\llbracket U+y\rrbracket),\im(\tau)\llbracket U+y\rrbracket\right)-2\im(\sigma)r\;\wedge \\[0.3cm]
&& \left.\ds x\leq\max\left(\im(\sigma)(r-\llbracket U+y\rrbracket),\im(\tau)\llbracket U+y\rrbracket\right)\right\}\,,
\end{array}
\eeq
and the poles in \eqref{rhopoles2}, and the following points lie below the contour
\beq
\begin{array}{rcl}
\ds\left\{-t_1-\ii\pi x\right.&\!\!\!\!|\!\!\!\!&\ds x\geq\min\left(\im(\sigma)\llbracket u_1+y\rrbracket,\,\im(\tau)(r-\llbracket u_1+y\rrbracket)\right)\,\wedge \\[0.3cm]
&& \left.\ds x\leq\max\left(\im(\sigma)\llbracket u_1+y\rrbracket,\,\im(\tau)(r-\llbracket u_1+y\rrbracket)\right)+2\im(\sigma)r\right\}\,,\\[0.3cm]
\ds\left\{A-2\ii\eta-\ii\pi x\right.&\!\!\!\!|\!\!\!\!&\ds x\geq\min\left(\im(\sigma)(r-\llbracket U-y\rrbracket),\,\im(\tau)(\llbracket U-y\rrbracket)\right)-2\im(\sigma)r\;\wedge \\[0.3cm]
&& \left.\ds x\leq\max\left(\im(\sigma)(r-\llbracket U-y\rrbracket),\,\im(\tau)(\llbracket U-y\rrbracket)\right)\right\}\,,
\end{array}
\eeq
These sets of points correspond to lines in the complex plane with constant real part. Depending on the values of $\re(t_1)$ and $\re(A)$, one should translate a set of points by $2\pi k$ if needed, so that the points always lie in the strip $0\leq\re z\leq2\pi$, and the contour of integration remains in this strip.  Using \eqref{Ifunct} one then performs $n$ shifts on the variable $t_1$, in the form $t_1\rightarrow t_1+\pi\tau r$, until $t_1+\pi\tau rn$ enters the set of points $\left\{t_1+\pi\sigma r x\,|\,1\leq x \leq 2\right\}$.  Then one transforms $t_1\rightarrow t_1-\pi\sigma r$. Under these transformations the poles of the integrand $\rho$ never cross the contour of integration.  Thus one obtains
\beq
I(t_1+\pi\tau rj-\pi\sigma rk,t_2,\ldots,t_5,u_1,\ldots,u_5)=I(t_1,\ldots,t_5,u_1,\ldots,u_5)\,,
\eeq
for all $j,k=0,1,\ldots$ such that $\pi\im(\tau)rj-\pi\im(\sigma)rk\in [0,\pi\im(\sigma)r]$.

The set of such points is dense thus $I$ does not depend on $t_1$, and by symmetry on any $t_i$.  Then from \eqref{Ifunct2} one has
\beq
I(t_1+\pi\sigma,t_2,\ldots,t_5,u_1-1,u_2,\ldots,u_5)=I(t_1,\ldots,t_5,u_1-1,u_2,\ldots,u_5)=I(t_1,\ldots,t_5,u_1,\ldots,u_5)\,.
\eeq
It follows that $I$ also does not depend on $u_1$ and by symmetry on any $u_i$.   Thus $I$ can only depend on $\p$ and $\q$.

Set each $u_i=0$.  In the limit $t_1+t_2\rightarrow0$, $\rho$ vanishes and the only contribution to the integral $I$, is from two finite residues coming from poles which cross the contour of integration for $y=0$.  Then evaluating the integral $I$ in this limit from its residues gives the right hand side of \eqref{str2}.

By analytic continuation one may then extend the domain of parameter values to that allowed by the contour between the endpoints $z=0,2\pi$.

\app{The identity \eqref{thtfunct}}
\label{subsec:todo}
In order to analyse the identity \eqref{thtfunct} at its poles, it is convenient to write the residues of both sides of \eqref{thtfunct}, in terms of the following theta function\footnote{This more resembles the standard theta function appearing in the literature {\it e.g.} \cite{Spiridonov-proofs,Rains-transformations}, with a change of variables. }
\beq
\label{oldtheta}
\theta(z\,|\,\tau)=(\EXP^{\ii z};\q^r)_\infty(\EXP^{-\ii z}\q^r;\q^r)_\infty\,.
\eeq
Then one finds that arguments of theta functions appearing on both sides of \eqref{thtfunct} differ by a simple shift $\pi\tau rk$, $k\in\mathbb{Z}$.  The theta function \eqref{oldtheta} obeys the following useful identity
\beq
\label{thtident}
\theta(z+\pi\tau rk\,|\,\tau)=\frac{\theta(z\,|\,\tau)}{(-\EXP^{\ii z})^k\,\EXP^{\ii\pi\tau rk(k-1)/2}},\quad k\in\mathbb{Z}\,.
\eeq
To show that \eqref{thtfunct} holds requires repeated use of this identity.

The right hand side of the identity \eqref{thtfunct} is chosen to have an equal set of poles and residues with the left hand side.  One wants to show that both sides of \eqref{thtfunct} define elliptic functions, and share the same sets of poles and residues.  Then by Liouville's theorem the difference of both sides is a constant, which may be found to be zero.  To do this, it should first be shown that no additional poles appear on right hand side of \eqref{thtfunct}.  Also it should be shown that both sides are invariant under the shift $z\rightarrow z+\pi\tau r$, and thus define elliptic functions of $z$.  

With the use of \eqref{thtident}, and identities in Appendix C, the calculations involved are straightforward and are summarised below.
\\\\
{\it No additional poles appear on right hand side of \eqref{thtfunct} :}  On the right hand side of \eqref{thtfunct}, there should be no poles appearing at the points $2z=\pi\tau\left(jr-\llbracket-2y\rrbracket\right)$, or equivalently at $2z=\pi\tau\left(jr+\llbracket2y\rrbracket\right)$, $j\in\mathbb{Z}$.  By using the identity \eqref{thtident}, at the poles $2z=\pi\tau\left(jr-\llbracket-2y\rrbracket\right)$ the residues of the two terms on the right hand side of \eqref{thtfunct} will differ by a factor
\beq
\exp\left(\ii\pi k_{m}+\ii\pi\sigma k_p+\ii\pi\tau k_q+Ak_A+\sum_{i=1}^5{t_i}k_{t_i}\right)\,,
\eeq
for some $k_{m},k_p,k_q,k_A,k_{t_i}\in\mathbb{R}$.  It can be shown that this factor is independent of the integer $j$, and is in fact unity, {\it i.e} $k_p,k_q,k_A,k_{t_i}=0$ and $k_{m}=0\mbox{ mod }2$.  Thus no additional poles appear on the right hand side of \eqref{thtfunct}.
\\\\
{\it Invariance under $z\rightarrow z+\pi\tau r$: }  One may use the relation \eqref{thtident} to show that both sides are invariant under $z\rightarrow z+\pi\tau r$.  This is the most straightforward property to check.
\\\\
{\it Difference of both sides of \eqref{thtfunct} is zero: } Evaluating both sides of \eqref{thtfunct} at the point $z=-t_1-\pi\tau\llbracket -u_1-y\rrbracket$, the left hand side gives -1, and the right hand side gives
\beq
\ds\left.-\frac{\EXP^{\ii t_1/r}\EXP^{\ii z/r}\EXP^{2\ii\pi\left(\llbracket y-u_1+1\rrbracket+\llbracket-2y-1\rrbracket\right)/r}\,\theta(t_1+A,u_1+U\,|\,\tau)\,\prod_{i=1}^5\theta(t_i-z,u_i-y\,|\,\tau)}{\theta(-2z,-2y\,|\,\tau)\,\theta(A-z,U-y\,|\,\tau)\,\prod_{i=2}^5\theta(t_1+t_i,u_1+u_i\,|\,\tau)}\right|_{z=-t_1-\pi\tau\llbracket -u_1-y\rrbracket}\,.
\eeq
After inspection, all theta functions in the above expression cancel up to an overall factor that makes the whole expression equal to -1.  Thus the difference of both sides of \eqref{thtfunct} is zero at the point $z=-t_1-\pi\tau\llbracket -u_1-y\rrbracket$.  Then since it has previously been shown that both sides are elliptic functions of $z$ sharing the same set of poles and conrresponding residues, by Liouville's theorem the difference of both sides is a constant, which must be zero.  Thus the identity \eqref{thtfunct} holds.

\app{Useful identities}
Here $m\in\mathbb{Z}$, $\llbracket m\rrbracket$ denotes $m\mbox{ mod }r$, and $\llbracket m\rrbracket_\pm$ denotes $\llbracket m\rrbracket\llbracket-m\rrbracket$.
\beq
\llbracket m \rrbracket_\pm=\llbracket -m\rrbracket_\pm
\eeq

\beq
\llbracket-m\rrbracket+\llbracket m-1\rrbracket=r-1
\eeq

\beq
\llbracket m-1\rrbracket_\pm-\llbracket m\rrbracket_\pm+2\llbracket-m\rrbracket=r-1
\eeq

\beq
\llbracket m+1\rrbracket_\pm-\llbracket m\rrbracket_\pm+2\llbracket m\rrbracket=r-1
\eeq

\beq
((2\llbracket m\rrbracket -r)\llbracket m\rrbracket_\pm-(2\llbracket m-1\rrbracket-r)\llbracket m-1\rrbracket_\pm=-(r-1)(r-2)-6\llbracket-m\rrbracket+6\llbracket m\rrbracket_\pm
\eeq

\beq
((2\llbracket m\rrbracket -r)\llbracket m\rrbracket_\pm-(2\llbracket m+1\rrbracket-r)\llbracket m+1\rrbracket_\pm=(r-1)(r-2)+6\llbracket m\rrbracket-6\llbracket -m\rrbracket_\pm
\eeq

\bibliography{total32}

\newcommand\oneletter[1]{#1}
\begin{thebibliography}{10}

\bibitem{Bax82}
Baxter, R.~J.
\newblock {\em Exactly {S}olved {M}odels in {S}tatistical {M}echanics}.
\newblock Academic, London, 1982.

\bibitem{AuY87}
Au-Yang, H., McCoy, B.~M., Perk, J. H.~H., Tang, S., and Yan, M.-L.
\newblock Commuting transfer matrices in the chiral Potts models: Solutions of
  star-triangle equations with genus $> 1$.
\newblock Phys. Lett. {\bf A123} (1987) 219--223.

\bibitem{Baxter:1987eq}
Baxter, R.~J., Perk, J. H.~H., and Au-Yang, H.
\newblock New solutions of the star triangle relations for the chiral {P}otts
  model.
\newblock Phys. Lett. {\bf A128} (1988) 138--142.

\bibitem{Zam-fish}
Zamolodchikov, A.~B.
\newblock ``{F}ishing-net'' diagrams as a completely integrable system.
\newblock Phys. Lett. B {\bf 97} (1980) 63--66.

\bibitem{FZ82}
Fateev, V.~A. and Zamolodchikov, A.~B.
\newblock Self-dual solutions of the star-triangle relations in
  {$Z\sb{N}$}-models.
\newblock Phys. Lett. A {\bf 92} (1982) 37--39.

\bibitem{Kashiwara:1986}
Kashiwara, M. and Miwa, T.
\newblock A class of elliptic solutions to the star-triangle relation.
\newblock Nucl. Phys. {\bf B275} (1986) 121--134.

\bibitem{FV95}
Volkov, A.~Y. and Faddeev, L.~D.
\newblock Yang-{B}axterization of the quantum dilogarithm.
\newblock Zapiski Nauchnykh Seminarov POMI {\bf 224} (1995) 146--154.
\newblock English translation: J. Math. Sci. {\bf 88} (1998) 202-207.

\bibitem{BMS07a}
Bazhanov, V.~V., Mangazeev, V.~V., and Sergeev, S.~M.
\newblock Faddeev-Volkov solution of the Yang-Baxter Equation and Discrete
  Conformal Symmetry.
\newblock Nucl. Phys. {\bf B784} (2007) 234--258.

\bibitem{BMS07b}
Bazhanov, V.~V., Mangazeev, V.~V., and Sergeev, S.~M.
\newblock Exact solution of the Faddeev-Volkov model.
\newblock Phys. Lett. A {\bf 372} (2008) 1547--1550.

\bibitem{Bazhanov:2010kz}
Bazhanov, V.~V. and Sergeev, S.~M.
\newblock {A Master solution of the quantum Yang-Baxter equation and classical
  discrete integrable equations}.
\newblock Adv. Theor. Math. Phys. {\bf 16} (2012) 65--95.

\bibitem{BKS2}
Bazhanov, V.~V., Kels, A.~P., and Sergeev, S.~M.
\newblock Quasi-classical expansion of the star-triangle relation and
  integrable systems on quad graphs.
\newblock In preparation  (2015).

\bibitem{Bax02rip}
Baxter, R.~J.
\newblock A rapidity-independent parameter in the star-triangle relation.
\newblock In {\em Math{P}hys {O}dyssey, 2001}, volume~23 of {\em Prog. Math.
  Phys.}, pages 49--63. Birkh\"auser Boston, Boston, MA, 2002.

\bibitem{Spiridonov-beta}
Spiridonov, V.~P.
\newblock On the elliptic beta function.
\newblock Uspekhi Mat. Nauk {\bf 56} (2001) 181--182.

\bibitem{Spiridonov-essays}
Spiridonov, V.~P.
\newblock Essays on the theory of elliptic hypergeometric functions.
\newblock Russ. Math. Surv. {\bf 63} (2008) 405.

\bibitem{BS11}
Bazhanov, V.~V. and Sergeev, S.~M.
\newblock Elliptic gamma-function and multi-spin solutions of the
  {Y}ang-{B}axter equation.
\newblock Nucl. Phys. {\bf B856} (2012) 475--496.

\bibitem{BKS}
Bazhanov, V.~V., Kels, A.~P., and Sergeev, S.~M.
\newblock Comment on star-star relations in statistical mechanics and elliptic
  gamma-function identities.
\newblock J. Phys. A: Math. Theor. {\bf 46} (2013) 152001.

\bibitem{DolanOsborn}
Dolan, F. and Osborn, H.
\newblock Applications of the superconformal index for protected operators and
  q-hypergeometric identities to dual theories.
\newblock Nucl. Phys. {\bf B818} (2009) 137--178.

\bibitem{Spiridonov-statmech}
Spiridonov, V.~P.
\newblock Elliptic beta integrals and solvable models of statistical mechanics.
\newblock Contemp. Math. {\bf 563} (2012) 181--211.

\bibitem{Yamazaki2012}
Yamazaki, M.
\newblock Quivers, {YBE} and 3-manifolds.
\newblock JHEP {\bf 1205} (2012) 147.

\bibitem{Yamazaki2013}
Yamazaki, M.
\newblock New Integrable Models from the Gauge/YBE Correspondence.
\newblock J. Stat. Phys. {\bf 154} (2014) 895--911.

\bibitem{K14}
Kels, A.~P.
\newblock A new solution of the star-triangle relation.
\newblock J. Phys. A: Math. Theor. {\bf 47} (2014) 055203.

\bibitem{Bax1}
Baxter, R.~J.
\newblock Solvable eight-vertex model on an arbitrary planar lattice.
\newblock Philos. Trans. Roy. Soc. London Ser. A {\bf 289} (1978) 315--346.

\bibitem{Bax72}
Baxter, R.~J.
\newblock Partition function of the eight-vertex lattice model.
\newblock Ann. Physics {\bf 70} (1972) 193--228.

\bibitem{Str79}
Stroganov, Y.~G.
\newblock A new calculation method for partition functions in some lattice
  models.
\newblock Phys. Lett. A {\bf 74} (1979) 116--118.

\bibitem{Zam79}
Zamolodchikov, A.~B.
\newblock {$Z\sb{4}$}-symmetric factorized {$S$}-matrix in two space-time
  dimensions.
\newblock Comm. Math. Phys. {\bf 69} (1979) 165--178.

\bibitem{Bax82inv}
Baxter, R.~J.
\newblock The inversion relation method for some two-dimensional exactly solved
  models in lattice statistics.
\newblock J. Stat. Phys. {\bf 28} (1982) 1--41.

\bibitem{Rui-EGF}
Ruijsenaars, S. N.~M.
\newblock First order analytic difference equations and integrable quantum
  systems.
\newblock J. Math. Phys. {\bf 38} (1997) 1069--1146.

\bibitem{Sp2008}
Spiridonov, V.~P.
\newblock Continuous biorthogonality of the elliptic hypergeometric function.
\newblock Algebra i Analiz {\bf 20} (2008) 155--185.
\newblock [English translation: St. Petersburg Math. J. {\bf 20} (2009)
  791--812].

\bibitem{GR13}
Gahramanov, I. and Rosengren, H.
\newblock A new pentagon identity for the tetrahedron index.
\newblock JHEP {\bf 2013} (2013) 128.

\bibitem{BNY13}
Benini, F., Nishioka, T., and Yamazaki, M.
\newblock 4d index to 3d index and 2d topological quantum field theory.
\newblock Phys. Rev. D {\bf 86} (2012) 065015.

\bibitem{Y14}
Yamazaki, M.
\newblock Four-dimensional superconformal index reloaded.
\newblock Theoretical and Mathematical Physics {\bf 174} (2013) 154--166.

\bibitem{AuYangPerk:ninfcp}
Au-Yang, H. and Perk, J.~H.
\newblock The large-N limits of the chiral Potts model.
\newblock Physica A {\bf 268} (1999) 175--206.

\bibitem{Rains-transformations}
Rains, E.~M.
\newblock Transformations of elliptic hypergeometric integrals.
\newblock Ann. Math. {\bf 171} (2010) 169--243.

\bibitem{GahSpi}
Gahramanov, I. and Spiridonov, V.~P.
\newblock The star-triangle relation and 3d superconformal indices.
\newblock arXiv:hep-th/1505.00765  (2015).

\bibitem{Spiridonov-proofs}
Spiridonov, V.~P.
\newblock Short proofs of the elliptic beta integrals.
\newblock Ramanujan J. {\bf 13} (2007) 265--283.

\bibitem{WilfZeil}
Wilf, H. and Zeilberger, D.
\newblock An algorithmic proof theory for hypergeometric (ordinary and “q”)
  multisum/integral identities.
\newblock Invent. Math. {\bf 108} (1992) 575--633.

\end{thebibliography}
\bibliographystyle{vvb-bibstyle}

\end{document}